\documentstyle[12pt,aps,prd,preprint,graphicx]{revtex}
\begin{document}
\title{The Inner Structure of  Black Holes}  
\author{Shahar Hod and Tsvi Piran}
\address{The Racah Institute for Physics, The
Hebrew University, Jerusalem 91904, Israel}
\date{\today}
\maketitle

\begin{abstract}
We study the gravitational collapse of a self-gravitating {\it
charged} scalar-field.  Starting with a {\it regular} spacetime, we
follow the evolution through the formation of an apparent horizon, a
Cauchy horizon and a final central singularity.  We find a null, {\it
weak}, mass-inflation singularity along the Cauchy horizon, which is a
precursor of a strong, {\it spacelike} singularity along the $r=0$
hypersurface. The inner black hole region is bounded (in the future)
by singularities. This resembles the classical inner structure of a
Schwarzschild black hole and it is remarkably different from the inner
structure of a charged static Reissner-Nordstr\"om or a stationary
rotating Kerr black holes.
\end{abstract}
\bigskip

The simple picture describing the exterior of a black-hole
\cite{Wheeler} is in dramatic contrast with its {\it interior}.  The
singularity theorems of Penrose and Hawking \cite{PenHaw} predicts the
occurrence of inevitable spacetime singularities as a result of a
gravitational collapse in which a black-hole forms.  According to the
{\it weak cosmic censorship} conjecture \cite{Penrose1}, these
spacetime singularities are hidden beneath the black-hole's
event-horizon.  However, these theorems tell us nothing about the
nature of these spacetime singularities. In particular, the {\it
  final} outcome of a generic gravitational collapse is still an open
question.

\begin{figure}
\begin{center}
\includegraphics[width=10cm]{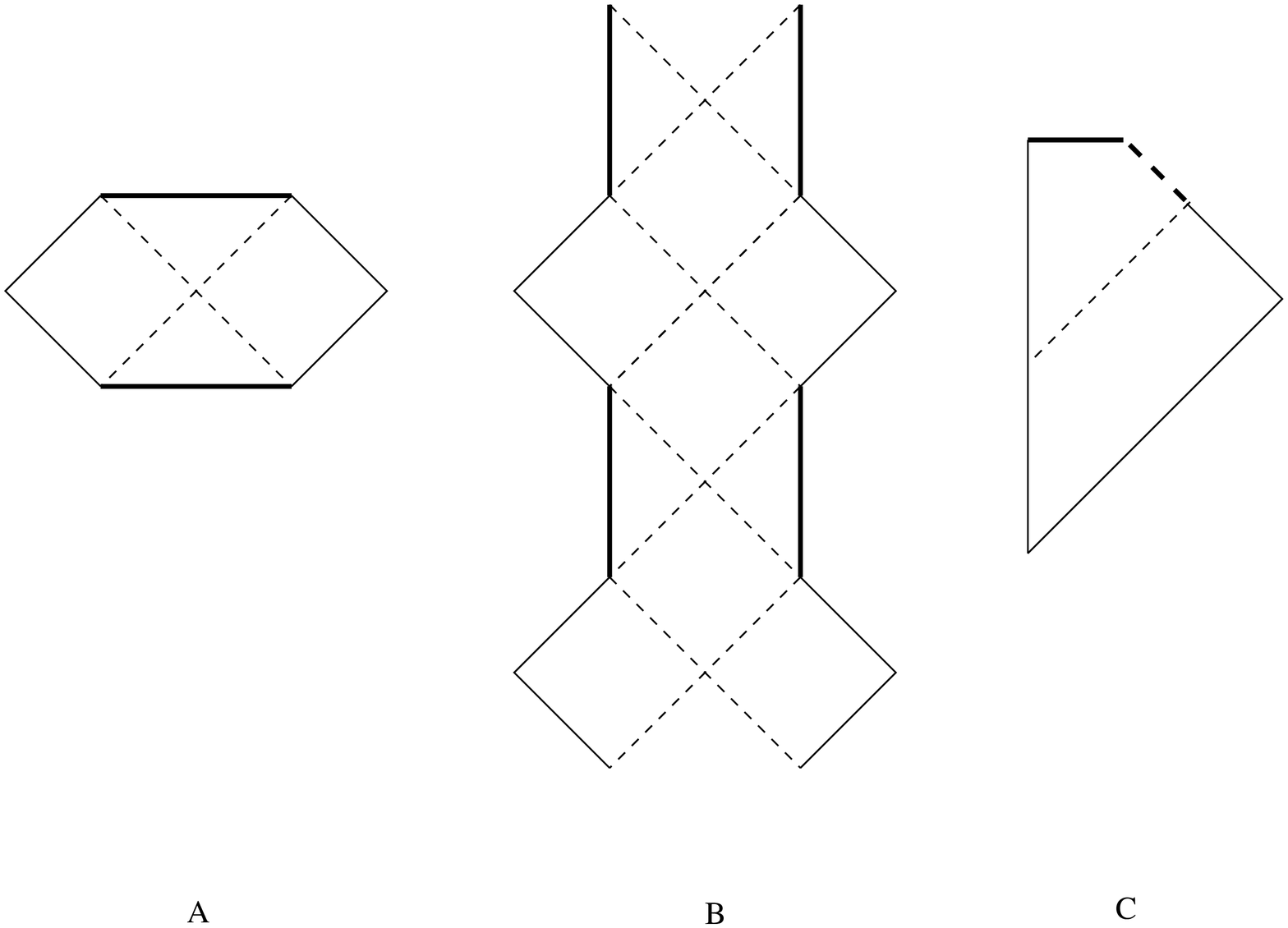}
\caption[Penrose Diagrams]{\label{conformal}
Penrose Diagrams of  Schwarzschild (a) Reissner-Nordstr\"om (b)
and Charged Collapse (c) spacetimes. 
Thick solid lines denote strong singularities, thick dashed lines
denote the weak null CH singularity, thin dashed lines denote the
various horizons.}
\end{center}
\end{figure}

Our physical intuition regarding the nature of these inner
singularities and the outcome of gravitational collapse is largely
based on the spherical Schwarzschild black hole solution and the
idealized Oppenheimer-Snyder collapse model \cite{Openheimer_snyder}.
The Schwarzschild black hole contains a strong {\it spacelike} central
singularity.  All the matter that falls into the black hole crashes
into this singularity within a finite proper time. The Schwarzschild
singularity is unavoidable.  This behaviour is manifested in the
Penrose diagram describing the conformal structure of a spacetime in
which a Schwarzschild black hole forms (see Fig. \ref{conformal}).

However, spherical collapse is not generic.  We expect some angular
momentum and this might change this picture drastically.  The inner
structure of a stationary rotating, Kerr, black hole contains a strong
inner {\it timelike} singularity, which is separated from external
observes by both an apparent horizon and a Cauchy horizon (CH). A
free-falling test particle cannot reach this singularity. Instead it
will cross a second Cauchy horizon and emerge form a white hole into
another asymptotically flat region. A remarkably similar structure
exists in a charged Reissner-Nordstr\"om black hole
(see Fig. \ref{conformal}) .  We do not
expect to find charged collapse in nature. However, this similarity
motivates us to study spherically symmetric charged gravitational
collapse as a simple toy model for a realistic generic rotating
collapse (which is at best axisymmetric).

Does the  inner structure of a Reissner-Nordstr\"om black hole
describe the generic outcome of gravitational collapse?  Novikov
\cite{Novikov} studied the collapse of a charged shell
and found that the shell will reach a minimal radius and bounce back,
emerging into another asymptotically flat region - a different
universe. The idea  of reaching other universes via a black hole's
interior is rather attractive.  It immediately captured the
imagination of the popular audience and  SciFi authors  coined
the ``technical'' term ``Stargate'' for this phenomenon. However as
predictability is lost at the CH this leads to  serious conceptual
problems.

We are faced with two gravitational collapse models. The ``traumatic''
collapse to Schwarzschild in which nothing can escape the central
singularity and the ``fascinating'' collapse to Kerr or
Reissner-Nordstr\"om in which a generic infalling observer might escape
unharmed to another Universe.  Which of the two possibilities is the
generic one?

Penrose, \cite{Penrose2} who was the first to address this
issue pointed out that small perturbations, which are remnants
of the gravitational collapse outside the collapsing object are
infinitely blueshifted as they propagate in the black-hole's interior
parallel to the Cauchy horizon.  The resulting infinite energy
leads to a curvature singularity. Matzner et. al
\cite{Matzner} have shown that the CH is indeed unstable to linear
perturbations.  This indicates that the CH might be singular -
``Stargate'' might be closed. A detailed modeling of this phenomena
suggests that the CH inside charged or spinning black-holes is
transformed into a {\it null}, {\it weak} singularity
\cite{Hiscock,Poisson,Ori1}.  The CH singularity is weak in the
sense that an infalling observer which hits this null singularity
experiences only a finite tidal deformation \cite{Ori1}.
Nevertheless, curvature scalars (namely, the Newman-Penrose Weyl
scalar $\Psi_{2}$) diverge along the CH, a phenomena known as {\it
  mass-inflation} \cite{Poisson}.

Despite this remarkable progress the physical picture is not complete
yet.  The evidence for the existence of a null, weak CH singularity is
mostly based on perturbative analysis.  The pioneering work of Gnedin
and Gnedin \cite{Gnedin} was a first step towards a full non-linear
analysis.  They have demonstrated the appearance of a central {\it
  spacelike} singularity deep inside a charged black-hole coupled to a
(neutral) scalar-field. Much insight was gained from the numerical
work of Brady and Smith \cite{Brady} who studied the same problem.
These authors established the existence of a {\it null} mass-inflation
singularity along the CH.  Furthermore, they showed that the singular
CH contracts to meet the central $r=0$ {\it spacelike} singularity.
More recently, Burko \cite{Burko} demonstrated that there is a good
agreement between the numerical results and the predictions of the
perturbative approach.

Still, the mass-inflation scenario has never been demonstrated
explicitly in a collapsing situation beginning from a regular
spacetime.  All previous numerical studies began with a singular
Reissner-Nordstr\"om spacetime with an additional infalling scalar
field.  We demonstrate here explicitly that mass-inflation
takes place during a dynamical charged gravitational collapse. We show
that the generic black hole that forms in a charged collapse is
engulfed by singularities in all future directions.

We consider the gravitational collapse of a self-gravitating {\it
charged} scalar-field. The physical model is described by the
coupled Einstein-Maxwell-charged scalar equations. We solve the
coupled equations using a characteristic method. Our scheme is based
on {\it double null} coordinates: 
%\begin{equation}\label{Eq1}
%ds^{2}=-\alpha(u,v)^{2}dudv+r(u,v)^{2}d\Omega ^{2}\  ,
%\end{equation}
a retarded null coordinate $u$ and an advanced null coordinate $v$.
The axis, $r=0$, is along $u=v$.  For $v \gg M$ our null ingoing
coordinate $v$ is proportional to the Eddington-Finkelstein null
ingoing coordinate $v_{e}$. These coordinates allow us to begin with a
regular initial spacetime (at approximately past null infinity),
calculate the formation of the black-hole's event horizon, and follow
the evolution inside the black-hole all the way to the central and the
CH singularities.

\begin{figure}
\begin{center}
{\bf a}\includegraphics[width=6cm]{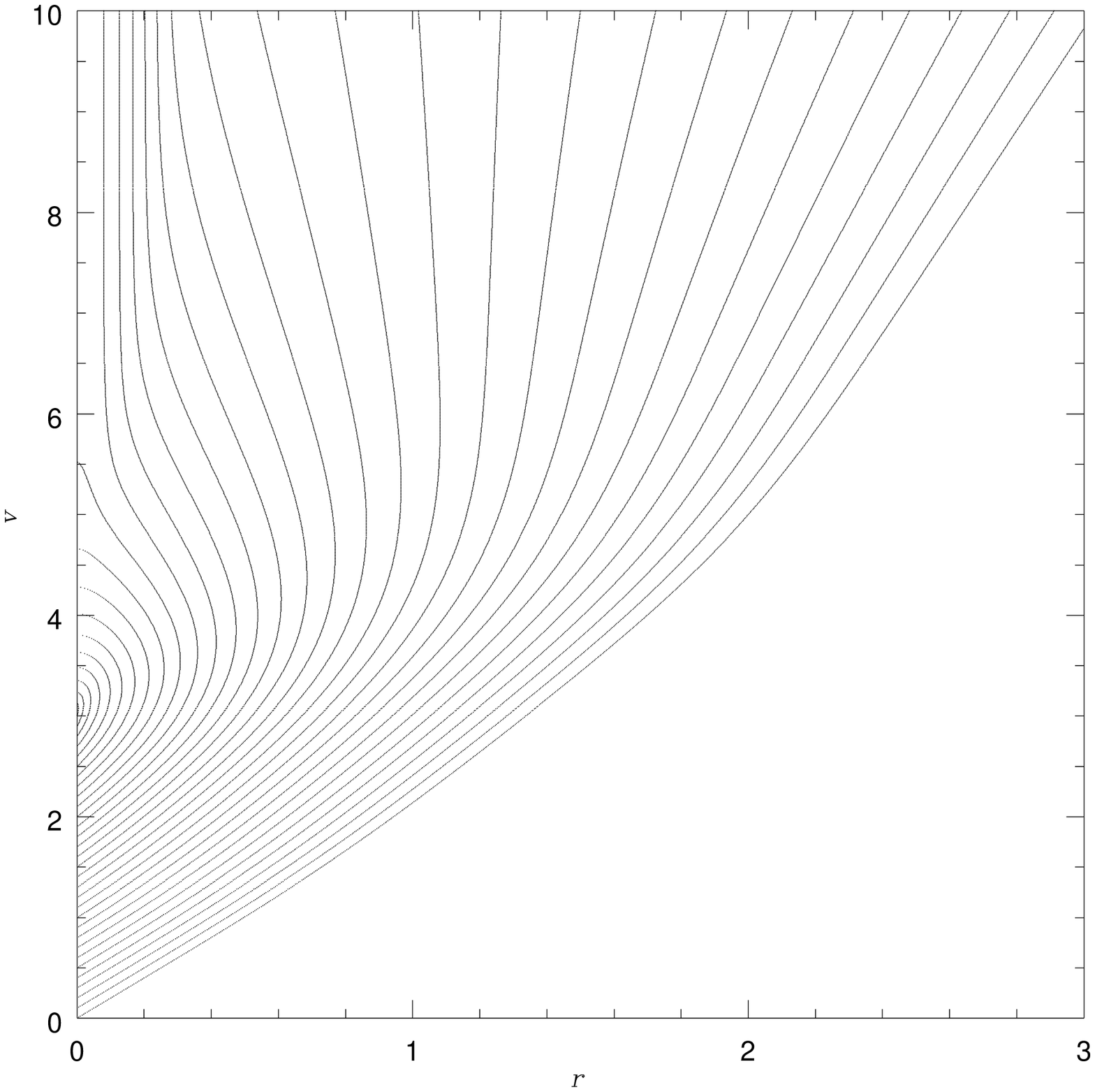}
{\bf b}\includegraphics[width=7.5cm]{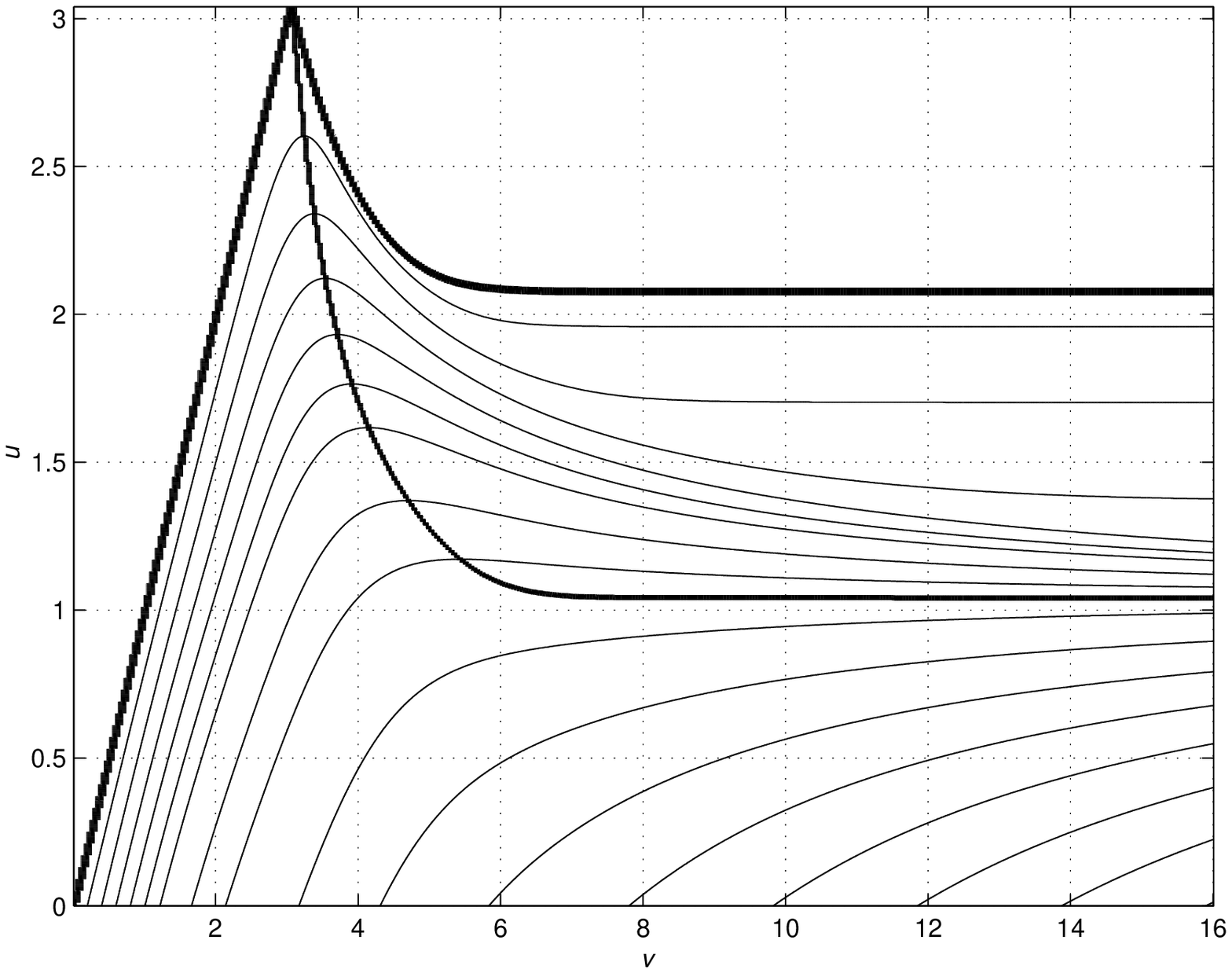}
\caption[The spacetime geometry]{\label{fig2}
\medskip
(a) Radial null rays originating from the regular  axis $r=0$.  The
outer-most rays escape to infinity, the inner-most rays terminate at
the singular section of the $r=0$ and the intermediate outgoing null
rays reach a ($u$-dependent) finite radius.
(b) Contour lines of the coordinate $r$ in the $vu$-plane.
The $r=0$ contour line is indicated by a thicker curve. Its left
section ($u=v$) is the {\it non-}singular axis, while its right
section corresponds the the central {\it spacelike} singularity. The
apparent horizon is indicated by $r_{v}=0$. The (singular) CH is a
null hypersurface located at $v \to \infty$. It contracts to meet
the central spacelike singularity (in a finite proper time).}  
\end{center}
\end{figure}

Fig. \ref{fig2} describes the numerical spacetime that we find.  The
upper panel (Fig. \ref{fig2}a) displays the radius $r(u,v)$ as a
function of the ingoing null coordinate $v$ along a sequence of
outgoing ($u=const$) null rays that originate from the {\it
  non}-singular axis $r=0$.  One can distinguish between {\it three}
types of outgoing null rays: (i) The outer-most (small-$u$) rays,
which {\it escape} to infinity.  (ii) The intermediate outgoing null
rays which approach a fixed radius $r_{CH}(u)$ at late-times $v \to
\infty$ indicating the existence of a CH.  (iii) The inner-most
(large-$u$) rays, which terminate at the {\it singular} section of the
$r=0$ hypersurface.  These outgoing rays reach the $r=0$ singularity
in a {\it finite} $v$, {\it without} intersecting the CH.  This
structure is drastically different from the Reissner-Nordstr\"om
spacetime, in which all outgoing null rays which originate inside the
black-hole intersect the CH.  Moreover while in a Reissner-Nordstr\"om
spacetime the CH is a {\it stationary} null hypersurface, here
$r_{CH}(u)$ depends on the outgoing null coordinate $u$. The CH {\it
  contracts} and reaches the inner $r=0$ singularity.  The CH is
smaller if the charge is smaller, and if the charge is sufficiently
small it is difficult (numerically) to notice the existence of a CH in the solution.

Fig. \ref{fig2}b. depicts the $r(u,v)$ contour lines. The outermost
contour line corresponds to $r=0$; its left section (a straight line
$u=v$) is the {\it non}-singular axis and its right section
corresponds to the central singularity at $r=0$. Since $r_{v}<0$ along
this section, the central singularity is {\it spacelike}.  Previously
$r_{v}=0$ indicated the existence of an apparent horizon (which is
first formed at $u \approx 1$ for this specific solution).  The CH is
a {\it null} hypersurface located at $v \to \infty$. This follows
because the intermediate outgoing null rays (in the range $1 \lesssim
u \lesssim 2.1$ for this specific solution) terminate at a finite
($u$-dependent) radius $r_{CH}(u)$.  The singular CH contracts to meet
the central ($r=0$) spacelike singularity (along the $u \simeq 2.1$
outgoing null ray).  Thus, the {\it null} CH singularity is a
precursor of the final {\it spacelike} singularity along the $r=0$
hypersurface.

\begin{figure}
\begin{center}
\includegraphics[width=12cm]{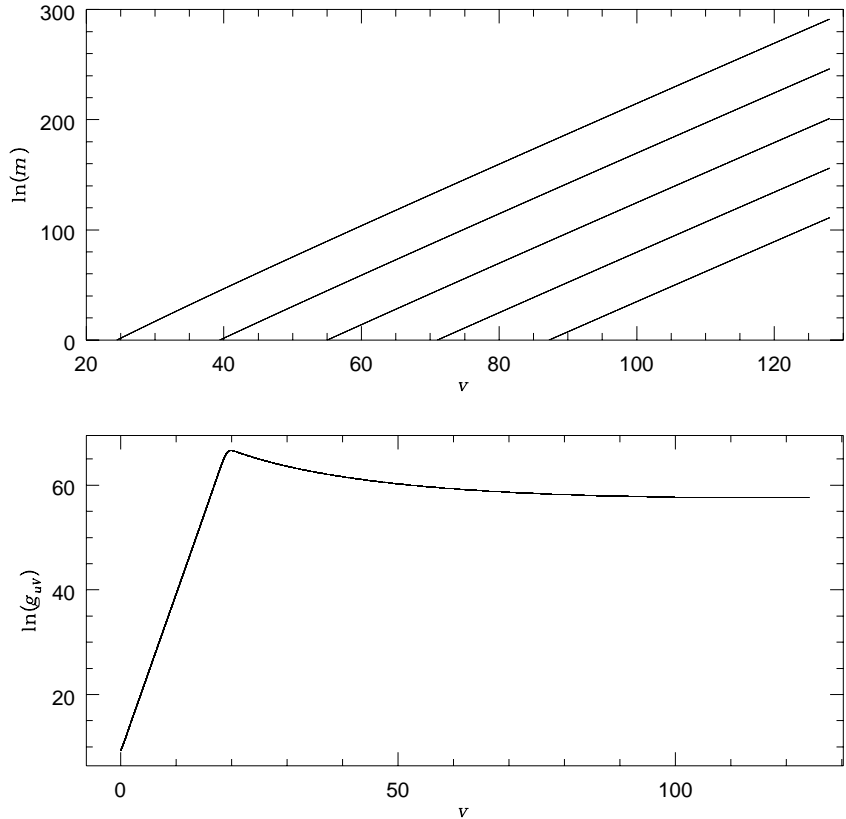}
\caption[mass-inflation]{\label{mass}
The CH singularity.  The top panel displays ln($m$) vs. advanced time
$v$, along a sequence of outgoing null rays. The exponential growth of
the mass-function demonstrates the appearance of the mass-inflation
scenario \cite{Poisson}.  The bottom panel displays the metric
function $g_{uV}$ along an outgoing null ray.  $V$ is a Kruskal-like
ingoing null coordinate. The CH is at $V \to 0$. $g_{uV}$ approaches a
  finite value as $V \to 0$ in agreement with the simplified model of
  Ori \cite{Ori1}. This demonstrates the {\it weakness} of the
  null mass-inflation singularity.}
\end{center}
\end{figure}

As expected from the {\it Mass Inflation} scenario the mass function
$m(u,v)$ (and the curvature) diverge exponentially along the outgoing
null rays (see Fig. \ref{mass}a).  The mass function increases not only
along the outgoing ($u$=const) null rays (as $v$ increases) but also
along ingoing ($v$=const) null rays (as $u$ increases).  The {\it
  weakness} of the singularity is demonstrated here by the metric
function $g_{uV}$ (see Fig.  \ref{mass}b) 
which approaches a finite value at the CH.   This
confirms the analytical analysis of Ori \cite{Ori1}, according to
which a suitable coordinate transformation can produce a {\it
  non-}singular metric.

Our numerical solution has put together all the different ingredients
found in the previous analyses into a single coherent picture.  The
inner structure of a  black hole that forms in a gravitational
collapse of a charged scalar-field is remarkably different from the
inner structure of a Reissner-Nordstr\"om (or Kerr) black hole (see
Fig. \ref{conformal}).  The inner region is bounded by singularities
in all future directions:  a {\it spacelike} singularity forms on $r=0$
and a {\it null} singularity forms along the CH, which contracts and
meets the {\it spacelike} singularity at $r=0$.  This structure is
much closer to the ``traditional'' Schwarzschild inner structure than
to the seemingly more generic Reissner-Nordstr\"om or Kerr structures.
However, while the {\it spacelike} singularity is {\it strong}, the
null singularity along the CH is {\it weak}.  Matter is able to cross
this singularity without being crushed by tidal forces. Thus, in spite
of this ``singular'' picture, ``Stargate'' might not be completely
closed after all (provided that the travelers are willing to suffer a
strong, yet finite distortion). These travelers will not have, of
course, the slightest idea what is expected for them beyond the CH.
The weakness of the CH singularity leaves open the question of
predictability beyond the CH.

\bigskip
\noindent
{\bf ACKNOWLEDGMENTS}
\bigskip

This research was supported by a grant from the Israel Science Foundation.
TP thanks W. Israel for helpful discussions.

\end{document}